\documentclass[journal,twocolumn]{IEEEtran}
\usepackage{amsmath,amsfonts}
\usepackage{algorithmic}
\usepackage{algorithm}
\usepackage{array}
\usepackage{amssymb}
\usepackage{multirow}
\usepackage{textcomp}
\usepackage{stfloats}
\usepackage[caption=false, font=footnotesize]{subfig}
\usepackage{url}
\usepackage{multirow}
\usepackage{verbatim}
\usepackage{graphicx}
\usepackage{cite}
\usepackage{bm}
\usepackage{makecell}
\usepackage{booktabs}
\usepackage{enumerate}
\usepackage{hyperref}
\hypersetup{hidelinks,
            colorlinks=true,
            allcolors=blue,
            pdfstartview=Fit,
            breaklinks=true}
\begin{document}

\title{CARNet: Channel-Adaptive Receiver Network for Robust NextG Communications}

\author{Chao Jiang, Zhuo Xu, and Yongli Yan
\thanks{The authors are with the Department of Electronic Engineering and the State Key Laboratory of Space Network and Communications, Tsinghua University, Beijing 100084, China (e-mail: jiangc24@mails.tsinghua.edu.cn; xz23@mails.tsinghua.edu.cn; yanyongli@tsinghua.edu.cn).}}

\maketitle
\begin{abstract}
Neural receivers have been recognized as a promising paradigm for the next-generation (NextG) communications.
However, due to the reliance on a static network optimized for specific channel conditions, their generalization capability across diverse scenarios remains a significant challenge.
To address this issue, this paper proposes a novel channel-adaptive neural receiver network (CARNet) based on the mixture-of-experts (MoE) framework.
The proposed architecture employs multiple expert networks together with an efficient routing mechanism to enable signal detection in various scenarios.
The experts are constructed via stacked ResNet blocks and specialize in robust signal detection within specific channel conditions, while the routing mechanism incorporates a lightweight representation learning module, which projects the coarse channel estimate into a low-dimensional latent embedding.
The learned embedding characterizes task-relevant channel conditions and provides efficient guidance for accurate expert selection.
Link-level simulation experiments demonstrate that the proposed CARNet achieves superior performance across diverse channel conditions.
\end{abstract}

\vspace{-15pt}
\section{Introduction} \label{Introduction}
Integration of artificial intelligence (AI) into physical layer design has recently been envisioned as a promising paradigm shift for the next-generation (NextG) communications~\cite{AI-Native-Air-Interface}.
To accelerate this paradigm transition, the 3rd Generation Partnership Project (3GPP) has officially advanced the standardization of AI-native air interfaces in Release 20 framework~\cite{3GPPRelease20}.

Among numerous AI-native air interface approaches, neural receivers have emerged as a promising enabler for NextG advanced receiver design.
Unlike conventional modular receivers that rely on simplified channel assumptions and individually design channel estimation, equalization, and demapping modules, neural receivers leverage deep neural networks to extract channel characteristics and jointly optimize these processing modules in a data-driven manner, demonstrating superior performances over traditional receivers in various scenarios~\cite{honkala2021deeprx}.

Recent studies on neural receivers have investigated various architectures to enhance performances in practical communication systems.
Specifically, the authors in~\cite{DARNet} propose a deep attention receiver network that directly processes time-domain waveforms to better capture channel characteristics and inter-symbol interference.
The research~\cite{CRNN-ResNet} combines convolutional neural networks and bidirectional long short-term memory to extract both temporal and frequency domain features for improved detection performances.
Moreover, to mitigate the non-linear distortions in transmission systems, \cite{DRNet} proposes a time-frequency dual-branch architecture to recover distorted signals.
To reduce pilot overhead and improve spectral efficiency, the authors in~\cite{CP-Pilot-Free, Superimposed-Pilot-Iterative} explore data transmission schemes without pilots or cyclic prefix.
Meanwhile, an energy-efficient neural receiver based on spiking neural networks is developed in~\cite{SNN-NeuralRx} with significantly reduced energy consumption.
In addition to the above data-driven models, hybrid neural receiver architectures are further proposed in~\cite{GNN-BiGAMP, BP-into-NN} by incorporating belief propagation and message passing into neural networks for enhanced interpretability and robustness.

Despite the remarkable advancements of the aforementioned neural receivers, achieving robust generalization across diverse scenarios remains a critical challenge.
Most existing schemes rely on a single set of fixed network parameters optimized for specific channel conditions.
However, in practical communication systems, user mobility and dynamic environment may lead to rapid variations in channel conditions.
Such static network is insufficient to accommodate diverse scenarios and thus suffers from performance degradation, particularly when confronted with unknown channel conditions that significantly deviate from the training distributions~\cite{AI-for-PHY-Survey}.
Therefore, overcoming the limitations of the single static network architecture to enhance robust generalization, has emerged as a fundamental challenge.

To overcome this issue, this paper introduces the advanced dynamic network architecture of the mixture-of-experts (MoE) incorporated with domain knowledge in wireless communications, and proposes a channel-adaptive neural receiver network (CARNet) for improved robustness and generalization.
Specifically, the main contributions can be summarized as follows:
\begin{itemize}
    \item We propose a channel-adaptive neural receiver architecture built upon the MoE framework, comprising a routing mechanism and multiple expert sub-networks.
    It enables dynamic expert selection and conditional computation to accommodate diverse channel conditions.
    \item To enhance routing efficiency and robustness, we design a lightweight representation learning module that extracts a compact latent embedding from the high-dimensional coarse channel estimate.
    This embedding is learned in a task-oriented manner and preserves channel-dependent statistical variations that are informative for distinguishing channel conditions and selecting appropriate experts.
    \item We conduct extensive link-level simulation experiments under various channel settings, including diverse channel models, signal-to-noise ratio (SNR) levels, delay spread, and Doppler shift.
    The simulation results demonstrate that the proposed scheme outperforms existing baselines and achieves superior generalization performances.
\end{itemize}

\section{System Model} \label{System Model}
\begin{figure*}
    \centering
    \includegraphics[width=0.85\linewidth]{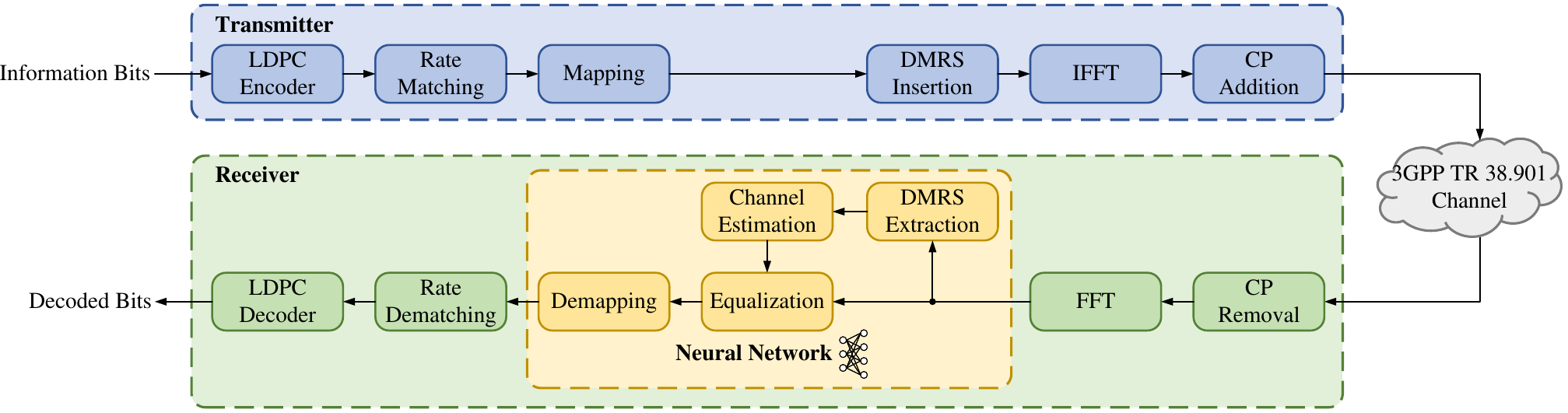}
    \caption{Overall block diagram of the 5G NR transceiver system.}
    \label{fig: 5G NR Transceiver}
\end{figure*}

The block diagram of the 5G New Radio (NR) transceiver system is shown in Fig.~\ref{fig: 5G NR Transceiver}.
In this paper, we consider a single-user uplink transmission scenario, where a single-antenna user equipment (UE) transmits information bits to a base station (BS) configured with $N_r$ receive antennas.

\subsection{5G NR Signal Transmission} \label{5G NR Signal Transmission}
As exhibited in Fig.~\ref{fig: 5G NR Transceiver}, the information bits are first encoded via low-density parity-check (LDPC) codes.
Following the rate matching procedure, the coded bits are mapped into complex-valued constellation symbols and then allocated onto the time-frequency resource grid, which comprises $S$ OFDM symbols and $F$ subcarriers.
To facilitate uplink channel estimation at the receiver side, demodulation reference signals (DMRS) are inserted into the specifically assigned resource elements (REs).
Finally, the frequency-domain OFDM symbols are converted into time-domain waveforms through the inverse fast Fourier transform (IFFT).
Before transmission, a cyclic prefix (CP) is prepended to each OFDM symbol to mitigate inter-symbol interference (ISI) induced by multipath propagation.

After propagating through multipath fading channels specified by standard 3GPP TR 38.901, the CP is removed at the BS and the fast Fourier transform (FFT) is then performed to obtain the frequency-domain signals.
The received signal at the $i$-th OFDM symbol ($i \in \{0, 1, \cdots, S-1\}$) and the $j$-th subcarrier ($j \in \{0, 1, \cdots, F-1\}$) can be given by
\begin{equation} \label{eqn: received signal}
    {\bf y}_{i,j} = {\bf h}_{i,j} x_{i,j} + {\bf n}_{i,j},
\end{equation}
where $x_{i,j} \in \mathbb{C}$ and ${\bf y}_{i,j} \in \mathbb{C}^{N_r \times 1}$ represent the transmitted and received symbols, respectively, ${\bf h}_{i,j} \in \mathbb{C}^{N_r \times 1}$ denotes the UE-BS channel matrix, and ${\bf n}_{i,j} \sim \mathcal{CN}({\bf 0}, \sigma_n^2 {\bf I}_{N_r})$ is the additive white Gaussian noise (AWGN) with variance $\sigma_n^2$.

\subsection{5G NR Receiver Processing} \label{5G NR Receiver Processing}
Conventional 5G NR receiver processing chain is illustrated in Fig.~\ref{fig: 5G NR Transceiver}.
After the CP removal and the FFT modules, the receiver extracts the DMRS symbols (i.e., pilots) to acquire the channel estimate.
According to~\eqref{eqn: received signal}, the least squares (LS) estimate can be computed as
\begin{equation} \label{eqn: LS estimate}
    {\hat{\bf h}}_{i,j} = {\bf y}_{i,j} p_{i,j}^*, \quad (i, j) \in \mathcal{P},
\end{equation}
where $p_{i,j}$ represents the known normalized pilot symbols satisfying $|p_{i,j}|^2 = 1$, and $\mathcal{P}$ denotes the index set of the REs allocated for pilots transmission.
A two-dimensional time-frequency interpolation is subsequently applied to acquire the complete channel estimate over the resource grid.
Based on the full channel estimate, the linear minimum mean square error (LMMSE) equalization is performed to reconstruct the transmitted symbols as
\begin{equation} \label{eqn: LMMSE equalization}
    {\hat x}_{i,j} = \left({\hat{\bf h}}_{i,j}^H {\hat{\bf h}}_{i,j} + \hat{\sigma}_n^2 {\bf I}\right)^{-1} {\hat{\bf h}}_{i,j}^H {\bf y}_{i,j}, \quad (i,j) \in \mathcal{D},
\end{equation}
where $\hat{\sigma}_n^2$ denotes the noise variance estimate obtained during the LS channel estimation phase, and $\mathcal{D}$ is the index set of the REs carrying data symbols.
Finally, the equalized symbols are converted into soft-bit information and passed to the LDPC decoder to recover the transmitted information bits.

\section{Proposed Channel-Adaptive Neural Receiver} \label{Proposed Channel-Adaptive Neural Receiver}
\subsection{Overview of the MoE-Based Neural Receiver} \label{Overview of the MoE-Based Neural Receiver}
Conventional neural receivers typically employ a deep neural network to jointly optimize channel estimation, equalization, and demapping modules in Fig.~\ref{fig: 5G NR Transceiver}.
However, existing approaches generally rely on single static network trained over specific channel conditions, leading to degraded generalization capability when deployed in diverse or unseen scenarios.

\begin{figure*}[!t]
    \centering
    \begin{minipage}[c]{0.54\textwidth}
        \centering
        \subfloat[Mechanism of the proposed CARNet.%
        \label{fig: CARNet}]{\includegraphics[width=\linewidth]{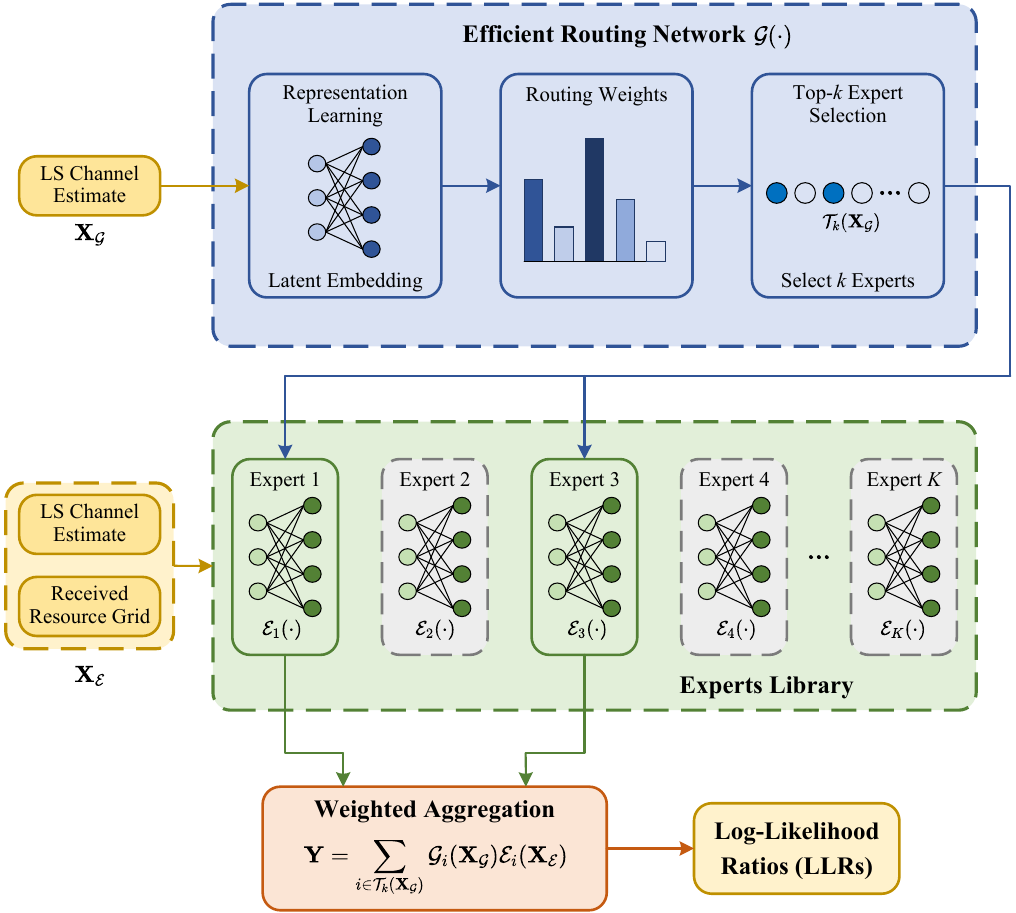}}
    \end{minipage}
    \begin{minipage}[c]{0.44\textwidth}
        \centering
        \subfloat[Detailed structure of the representation learning network.
        \label{fig: representation learning network}]{\includegraphics[width=\linewidth]{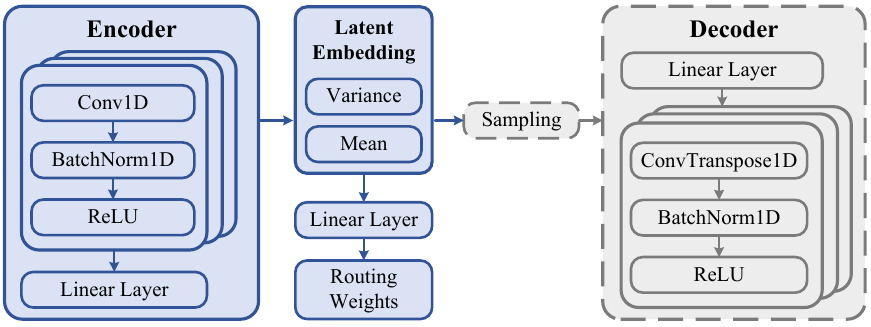}}
        \vspace{0.3cm}
        \subfloat[Detailed structure of the expert network.%
        \label{fig: expert network}]{\includegraphics[width=\linewidth]{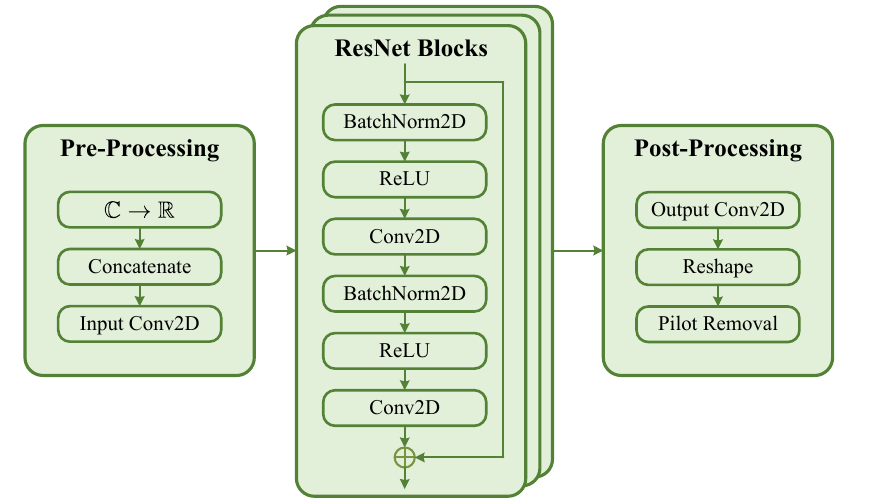}}
    \end{minipage}
    \caption{Overall architecture and detailed sub-network structures of the proposed CARNet.}
    \label{fig: overall architecture}
\end{figure*}

To overcome this limitation, this paper proposes a channel-adaptive neural receiver network based on the MoE framework, as illustrated in Fig.~\ref{fig: CARNet}.
Instead of employing single shared network, the proposed receiver comprises a lightweight routing network represented by $\mathcal{G}(\cdot)$, and a library of $K$ specialized homogeneous expert networks $\mathcal{E} = \{\mathcal{E}_1, \mathcal{E}_2, \dots, \mathcal{E}_K\}$.
By leveraging a divide-and-conquer strategy, the receiver can dynamically select suitable experts according to the current channel condition and thereby enables adaptive signal processing across different scenarios.
Mathematically, the final output, i.e., the log-likelihood ratios (LLRs), can be computed as the linear weighted combination of the outputs from the activated top-$k$ experts
\begin{equation} \label{eqn: weighted output}
    \text{LLRs} = \sum_{i \in \mathcal{T}_k({\bf X}_\mathcal{G})} \mathcal{G}_i({\bf X}_\mathcal{G}) \mathcal{E}_i({\bf X}_\mathcal{E}),
\end{equation}
where ${\bf X}_\mathcal{G}$ and ${\bf X}_\mathcal{E}$ denote the input of the routing and expert networks, respectively, $\mathcal{T}_k({\bf X}_\mathcal{G})$ is the index set of the top-$k$ experts assigned with the highest weights by the routing network, $\mathcal{E}_i({\bf X}_\mathcal{E})$ is the output of the $i$-th expert network, and $\mathcal{G}_i({\bf X}_\mathcal{G})$ denotes the dynamic routing weight corresponding to the $i$-th expert, satisfying $\mathcal{G}_i({\bf X}_\mathcal{G}) \geq 0$ and $\sum_{i \in \mathcal{T}_k({\bf X}_\mathcal{G})} \mathcal{G}_i({\bf X}_\mathcal{G}) = 1$.

\subsection{Efficient Routing via Latent Representation}
The primary objective of the routing network is to identify the current channel condition and then determine which expert networks should be activated.
However, a generic data-driven routing strategy without integrating any delicate design may struggle to distinguish intrinsic propagation characteristics and generalize across diverse scenarios.
Moreover, the corresponding routing decision may also be sensitive to noise levels and local estimate errors, leading to unstable expert selection.

Different from typically adopted MoE architectures, where the routing decisions are generally inferred from raw or high-dimensional features, the proposed routing network incorporates a representation learning module tailored to the structural properties of wireless channels.
Specifically, although the channels are represented as a high-dimensional matrix whose size scales with the numbers of transceiving antennas, OFDM symbols, and subcarriers, its intrinsic variations can typically be dominated by a limited number of physical propagation factors, such as angles of arrival, angles of departure, path loss, delay spreads, and Doppler shifts.
Additionally, due to the time-frequency-space coherence of the wireless channels, the channel coefficients of adjacent REs and receive antennas generally exhibit strong statistical correlations.
Therefore, the channels are inherently redundant and can be characterized by a low-dimensional latent representation.
Notably, this latent representation is not designed to explicitly estimate individual physical propagation parameters.
Instead, it serves as a low-dimensional and task-oriented representation that preserves channel-dependent statistical patterns for expert selection.

As illustrated in Fig.~\ref{fig: representation learning network}, the proposed representation learning network adopts variational autoencoder (VAE) structure to extract the latent representation from the LS channel estimate.
The channel estimate at the pilot positions is first reshaped into a one-dimensional sequence and processed through multiple stacked one-dimensional convolutional blocks, where the number of feature channels gradually increases while the length of feature maps progressively decreases.
This hierarchical design can compresses redundant features while preserving sufficient learning capacity to capture multi-scale propagation patterns.
The encoded feature maps are then flattened and projected into two latent statistics via a linear layer, namely the mean vector $\boldsymbol{\mu} \in \mathbb{R}^{D_z}$ and the log-variance vector $\log \boldsymbol{\sigma}^2 \in \mathbb{R}^{D_z}$
\begin{equation} \label{eqn: output of encoder}
    \boldsymbol{\mu}, \log \boldsymbol{\sigma}^2 = \text{Encoder}({\bf X}_\mathcal{G}),
\end{equation}
where $D_z$ denotes the dimension of the latent representation, $\boldsymbol{\mu}$ characterizes the deterministic latent embedding of the current propagation condition, while $\log \boldsymbol{\sigma}^2$ quantifies the uncertainty of the latent representation.
These two latent statistics jointly parameterize the compact probabilistic representation of the current channel conditions.

During training, the latent variable is generated through the reparameterization operation to maintain end-to-end differentiability, which can be formulated as
\begin{equation} \label{eqn: reparameterization}
    {\bf z} = \boldsymbol{\mu} + \boldsymbol{\epsilon} \odot \exp \left(\log \boldsymbol{\sigma}^2 / 2\right),
\end{equation}
where ${\bf z}$ denotes the sampled latent representation, $\odot$ is the element-wise product, and $\boldsymbol{\epsilon} \sim \mathcal{N}({\bf 0}, {\bf I}_{})$ represents a random sampling from the standard Gaussian distribution.
This sampling process encourages the encoder to describe the channel condition through a smooth probabilistic representation, thus the learned latent representation becomes less sensitive to local estimation errors and noise.
The decoder, which is constructed as a symmetric counterpart of the encoder, is then employed to recover the input from $\bf z$
\begin{equation} \label{eqn: output of decoder}
    {\widehat{\bf X}}_\mathcal{G} = \text{Decoder}({\bf z}).
\end{equation}

In the inference stage, the sampling module and the decoder are removed.
The latent embedding is fed into a multilayer perceptron (MLP) to generate routing weights $\boldsymbol{w} \in \mathbb{R}^K$ as
\begin{equation} \label{eqn: routing weights}
    \boldsymbol{w}= \text{Softmax}\Big(\text{MLP}\big(\text{Cat}[\boldsymbol{\mu}, \log \boldsymbol{\sigma}^2]\big) \Big).
\end{equation}
The routing network selects $k$ experts with the largest weights and the weights are normalized over these activated experts.

\vspace{-2pt}
\subsection{Specialized Expert Networks}
Upon receiving the routing weights from the well-designed routing network, the activated top-$k$ experts perform the actual signal recovery.
In the proposed CARNet, the expert library $\mathcal{E}$ consists of $K$ structurally homogeneous neural networks with different parameters.
This design can guarantee computational balance, while allowing each expert to implicitly specialize in distinct channel conditions identified by the routing network.

As illustrated in Fig.~\ref{fig: expert network}, each expert network is built upon a fully convolutional residual network (ResNet) backbone.
The network takes two components as inputs: the received signal ${\bf Y} \in \mathbb{C}^{N_r \times S \times F}$ and the interpolated LS channel estimate ${\widehat{\bf H}} \in \mathbb{C}^{N_r \times S \times F}$.
Notably, the latter provides a bootstrap for channel estimation and can facilitate training process.
Each expert first employs a convolutional layer to project the received signal into a multi-scale feature space.
The resulting feature maps are then processed by $N_B$ stacked ResNet blocks for deep feature extraction.
Each block consists of two convolutional layers with batch normalization and nonlinear activation, together with a residual shortcut.
Finally, an output convolutional layer integrates extracted features to bit-wise LLRs.
After masking the output LLRs at the pilot positions, the remaining LLRs corresponding to the data REs are rearranged and flattened.

Although all experts share a homogeneous architecture, they are encouraged to learn different signal processing behaviors through channel-adaptive routing.
Since different latent representations activate different expert subsets, each expert tends to specialize in certain channel conditions.

\vspace{-2pt}
\subsection{Training Strategy}
End-to-end joint training from scratch is highly unstable.
It may suffer from the dilemma of expert collapse, where the routing network excessively favors a single expert, leaving the remaining experts unoptimized and the network degenerating back to a single-model receiver.
To ensure stable training, we propose a two-stage training strategy tailored for the CARNet.

In the first stage, the VAE-based routing network is pre-trained to achieve robust representation learning.
The module is optimized by minimizing the evidence lower bound (ELBO), comprising reconstruction loss and KL-divergence
\begin{equation} \label{eqn: training loss of routing network}
    \mathcal{L}_\mathcal{G} = \mathbb{E}\|\mathbf{X}_{\mathcal{G}} - \hat{\mathbf{X}}_{\mathcal{G}}\|^2 + \beta\mathbb{E}\left\{\sum_{d=1}^{D_z}\frac{\mu_d^2+\sigma_d^2-\log\sigma_d^2-1}{2}\right\}
\end{equation}
where $\beta$ is the regularization factor, and $\mathbb{E}(\cdot)$ is approximated by the average over the training batch.
 
In the second stage, the pretrained encoder is incorporated into the routing network and jointly optimized with the expert networks.
To preserve the channel representation learned during the first stage while adapting to the expert optimization, the encoder is fine-tuned with a much smaller learning rate than the expert networks.
For the signal detection task of the experts, we aim to minimize the binary cross entropy loss
\begin{equation}
    \mathcal{L}_{\mathrm{det}} = -\mathbb{E}\Big[b\log \hat{p}(b=1) + (1-b)\log \big(1-\hat{p}(b=1)\big)\Big],
\end{equation}
where $b$ represents the transmitted coded bits, and the estimated bit probability $\hat{p}(b=1)$ is obtained from the output LLRs through the Sigmoid function.
To avoid expert collapse, we introduce an auxiliary load balancing loss denoted by
\begin{equation}
    \mathcal{L}_{\mathrm{bal}} = \sum_{i=1}^{K} \left(\mathbb{E}(f_i)-\frac{k}{K}\right)^2,
\end{equation}
where $f_i$ denotes the activation frequency of the \(i\)-th expert.
Minimizing $\mathcal{L}_{\text{bal}}$ encourages more balanced expert utilization and ensures that all experts receive sufficient training samples.
The overall loss for joint training during the second stage is
\begin{equation}
    \mathcal{L}_{\mathrm{total}} = \mathcal{L}_{\mathrm{det}} + \alpha \mathcal{L}_{\mathrm{bal}},
\end{equation}
where $\alpha$ denotes the weighting coefficient of load balancing.

\section{Performance Evaluation} \label{Performance Evaluation}
\subsection{Simulation Setup} \label{Simulation Setup}
In this paper, we leverage NVIDIA Sionna library~\cite{hoydis2023sionna}, a typically adopted link-level simulation platform, to implement the 5G NR uplink transmission described in Section~\ref{System Model}.
The communication system configuration is summarized in Table~\ref{tab: Simulation Parameters}.

\begin{table}[!h]
    \vspace{-6pt}
    \caption{Simulation Parameters}
    \vspace{-3pt}
    \label{tab: Simulation Parameters}
    \centering
    \begin{tabular}{cc}
        \Xhline{0.8pt}
        \textbf{Parameter} & \textbf{Value} \\
        \Xhline{0.8pt}
        Carrier frequency & 3.5 GHz \\
        Number of subcarriers & 256 \\
        Subcarrier spacing & 15 kHz \\
        Number of OFDM symbols & 14 (1 slot) \\
        Channel model & CDL-A/B/C/D/E, UMi, UMa \\
        RMS delay spread & $10$ ns $\sim 300$ ns \\
        UE speed & $0$ m/s $\sim 35$ m/s \\
        Number of BS antennas & 2 \\
        Number of bits per symbol & 4 (16-QAM) \\
        \Xhline{0.8pt}
    \end{tabular}
\end{table}

Notably, the proposed CARNet is configured with $K = 3$ experts and each expert contains $N_B = 4$ ResNet blocks.
Each ResNet block comprising two Conv2D layers with 128 output channels and a $3 \times 3$ kernel size.
The lightweight encoder of the representation learning module contains three Conv1D layers, where the output channels are set to 16, 32, and 64, with a kernel size of 3 and a stride of 2.
In this work, the routing network activates top $k = 1$ expert for each input.

For performance comparison, multiple baselines are considered:
5G NR receiver LS-LMMSE that perfroms LS channel estimation and LMMSE equalization, DeepRx~\cite{honkala2021deeprx}, CRNN-ResNet~\cite{CRNN-ResNet}, and perfect CSI with LMMSE equalization.
To verify the effectiveness of the representation learning module, the CARNet-Base removes this module and leverage a deep neural network to generate routing weights.
To investigate the performances across diverse channel conditions, the simulation cover different channel models (CDL-B/C/D and UMa for training, with CDL-A/E and UMi as unseen scenarios), as well as varying SNR, delay spreads, and UE speeds in Table~\ref{tab: Simulation Parameters}.

\subsection{Simulation Results} \label{Simulation Results}
\begin{figure*}[!t]
    \vspace{-5pt}
    \centering
    \subfloat[CDL-C.\label{fig: CDL-C}]{\includegraphics[width=0.33\textwidth]{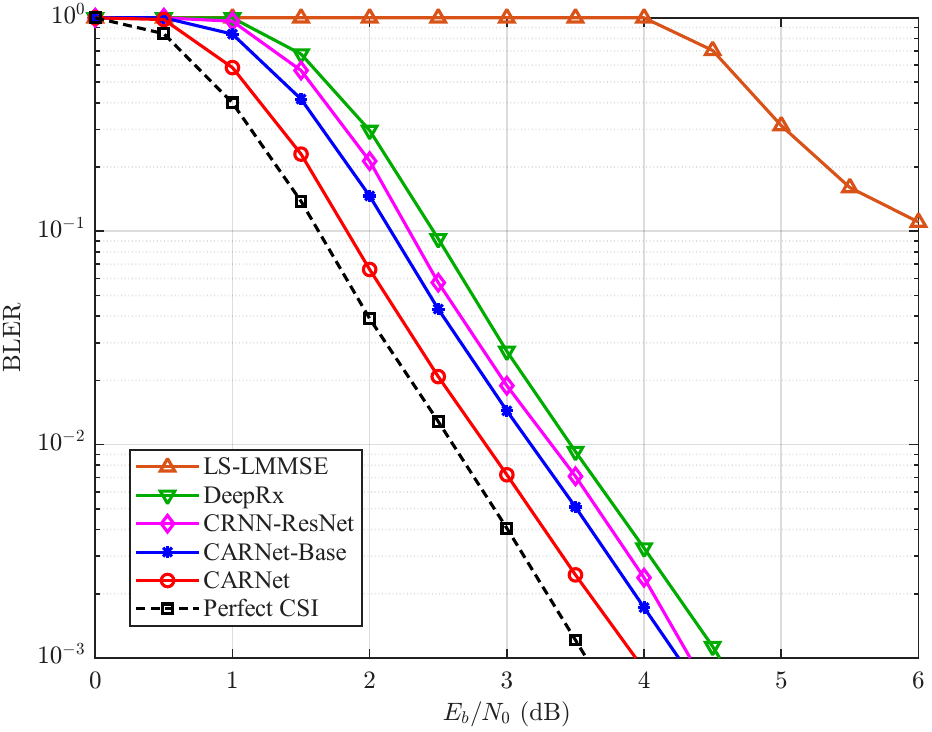}}
    \subfloat[CDL-D.\label{fig: CDL-D}]{\includegraphics[width=0.33\textwidth]{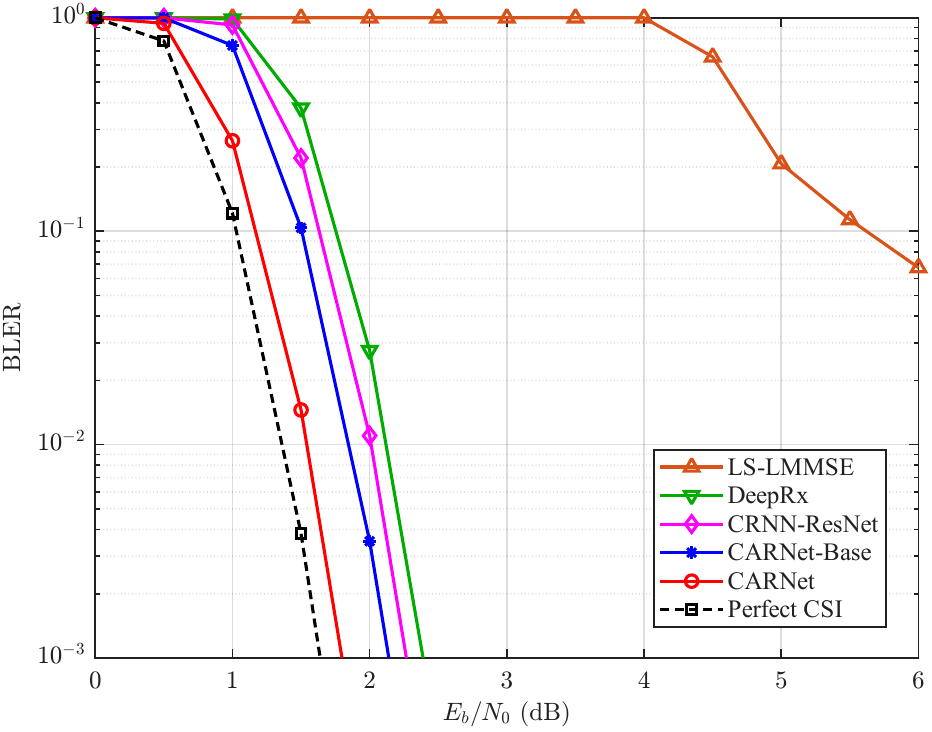}}
    \subfloat[UMi.\label{fig: UMi}]{\includegraphics[width=0.33\textwidth]{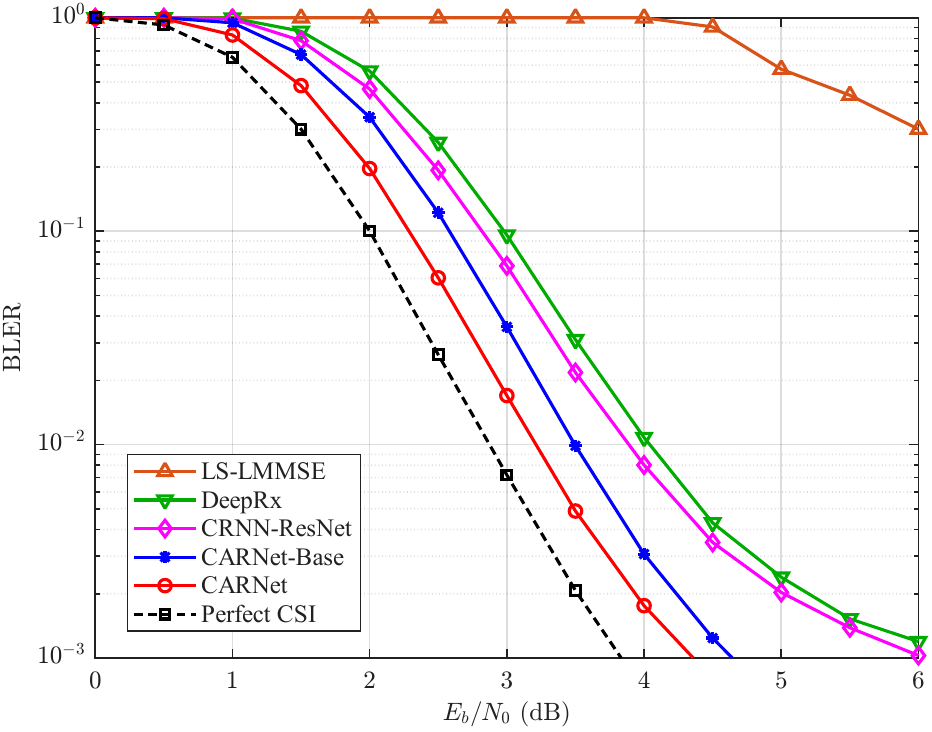}}
    \caption{BLER performance versus $E_b/N_0$ under diverse channel conditions: (a) NLoS CDL-C, (b) LoS CDL-D ,and zero-shot generalization to (c) UMi.}
    \vspace{-7pt}
    \label{fig: BLER performance}
\end{figure*}

As illustrated in Fig.~\ref{fig: CDL-C} and Fig.~\ref{fig: CDL-D}, we first evaluate the block error rate (BLER) performance under CDL-C and CDL-D, which are included in the training dataset.
The former represents a NLoS scenario with rich multipath propagation, while the latter is LoS-dominant environment.
The proposed CARNet outperforms static neural receiver baselines, including DeepRx
and CRNN-ResNet, and approaches the Perfect-CSI benchmark, indicating that CARNet can adapt to different channel conditions by dynamically selecting suitable experts.

Furthermore, Fig.~\ref{fig: UMi} evaluates the zero-shot generalization performance under the unseen UMi scenario.
The proposed CARNet still achieves lower BLER than DeepRx and CRNN-ResNet.
The performance gap becomes more significant in the high $E_b/N_0$ region, due to model mismatch and insufficient adaptation to unseen propagation conditions.
The results demonstrate that the proposed CARNet can maintain superior performance under unseen channel distributions.

To verify the effectiveness of the proposed representation learning module, we additionally conduct an ablation study.
As observed from Fig.~\ref{fig: BLER performance}, the CARNet consistently outperforms the CARNet-Base without the delicately designed representation learning module.
This indicates that directly inferring routing weights from data-driven features can be less effective for channel-adaptive expert selection.
In contrast, the proposed representation learning module extracts a compact and task-oriented channel embedding, which provides more reliable guidance for accurate expert selection.

\begin{table}[htbp]
    \vspace{-10pt}
    \caption{Complexity Comparison}
    \vspace{-5pt}
    \label{tab: Complexity Comparison}
    \centering
    \begin{tabular}{ccc}
        \Xhline{0.8pt}
        \textbf{Neural Receiver} & \textbf{Params (M)} & \textbf{GFLOPs} \\
        \Xhline{0.8pt}
        DeepRx & 1.196 & 4.158 \\
        CRNN-ResNet & 1.273 & 5.461 \\
        CARNet-Base & 3.931 & 4.637 \\
        CARNet & 3.810 & 4.218 \\
        \Xhline{0.8pt}
    \end{tabular}
\end{table}

Table~\ref{tab: Complexity Comparison} compares the complexity of different neural receivers, including the number of parameters and the floating-point operations (FLOPs) per slot.
Although CARNet has more parameters due to the expert library, its GFLOPs remain moderate owing to sparse conditional computation.
Therefore, the inference complexity of CARNet is close to DeepRx and lower than CRNN-ResNet.
Moreover, compared with CARNet-Base, CARNet achieves reductions in both parameters and GFLOPs, indicating that the representation learning module provides a more computationally efficient routing mechanism.

\section{Conclusion} \label{Conclusion}
This paper introduces a channel-adaptive neural receiver network (CARNet) for improved robustness and generalization.
CARNet extracts task-oriented channel embeddings via a delicately designed representation learning module and dynamically selects specialized experts for signal detection under current scenarios.
Simulation results demonstrate that CARNet outperforms existing receivers across diverse conditions.

\bibliographystyle{IEEEtran}
\bibliography{references.bib}

\end{document}